\documentclass[12pt]{article}
\usepackage{amssymb}
\usepackage{amsmath}
\usepackage{color}
\newcommand{\bea}{\begin{eqnarray}}
\newcommand{\eea}{\end{eqnarray}}
\newcommand{\be}{\begin{equation}}
\newcommand{\ee}{\end{equation}}
\newcommand{\beann}{\begin{eqnarray*}}
\newcommand{\eeann}{\end{eqnarray*}}
\newcommand{\x}{{\bf x}}
\newcommand{\R}{{\mathbb R}}

\def\E{{\mathbb E}}

\newtheorem{lemma}{Lemma}

\title{Bandwidth selection for kernel estimators \\
       of the spatial intensity function}
\author{
O. Cronie$^{\dagger}$\footnote{Correspondence: ottmar.cronie@umu.se, Marie-Colette.van.Lieshout@cwi.nl} 
and 
M.N.M.\ van Lieshout$^{\ddagger *}$
}
\date{}

\begin{document}
\newcommand{\edt}[1]{{\vbox{ \hbox{#1} \vskip-0.3em \hrule}}}

\maketitle \noindent 
\begin{center} 
$^{\dagger}$ Department of Mathematics and Mathematical Statistics, \\
Ume{\aa} University, 
901 87 Ume\aa, Sweden
\\
$^{\ddagger}$ CWI, 
P.O. Box 94079, 
1090 GB Amsterdam, 
The Netherlands \\
Department of Applied Mathematics,
University of Twente, \\
P.O. Box 217, 
7500 AE Enschede, 
The Netherlands
\end{center} 

\bigskip

\noindent{\bf Abstract}: 
We discuss and compare various approaches to the problem of bandwidth 
selection for kernel estimators of intensity functions of spatial point 
processes. We also propose a new method based on the Campbell formula applied to 
the reciprocal intensity function. The new method is fully non-parametric, 
does not require knowledge of the product densities, and is not restricted 
to a specific class of point process models. 

\


\


\section{Introduction}
\label{S:intro}

Spatial point patterns arise in many scientific domains. Typical examples 
include the map of trees in a forest stand, the addresses of individuals 
infected with some disease or the locations of cells in a tissue (see 
e.g.\ \cite{Digg14,Gelf10,Illi08}). 

The analysis of such patterns usually includes estimating the intensity
function, that is, the likelihood of finding a point as a function of
location. Sometimes the scientific context suggests a parametric form 
for the intensity function, perhaps in terms of covariate information. 
More often, non-parametric estimation is called for. In both cases,
it is important that one is able to estimate the intensity function 
in a reliable way. Indeed, this is more urgent than ever in light of 
the recent development of functional summary statistics that correct 
for spatial heterogeneity \cite{Badd00,Lies11} and the current paper
was motivated by our recent work \cite{CronLies16} in this direction.

Here we focus on non-parametric estimation. A few approaches have been 
suggested in the literature. For example, one may divide the observation 
window into quadrats and count the number of points that fall in each. 
An obvious drawback of this method is its strong dependence on the size 
and shape of the quadrats. A partial solution is to use spline
\cite{Ogat98} or kernel smoothing, where the shape of the kernel used 
tends to have little impact; the size parameter, though, remains crucial. 
Therefore, data dependent methods have been proposed in which the quadrants 
are replaced by the cells in the Delaunay \cite{Scha07,SchaWeyg00} or 
Voronoi tessellation \cite{BernWeyg96,Ord78} of the pattern. The relative 
merits of these approaches were investigated by \cite{BarrScho10,Lies12}. 
Further details can be found in \cite[Chapters~3 and 5]{Digg14}.

The goal of this paper is to discuss and compare various approaches to 
the problem of choosing the size parameter, the {\em bandwidth\/},
when performing kernel estimation. The bandwidth is the parameter determining to what degree the variations in the intensity are 
smoothed out. At first glance, this might appear to be a trivial task.
It turns out to be extremely challenging, though. Indeed, some authors 
argue that the problem is unsolvable \cite{Davi14}. Apart from various rules
of thumb (see \cite{Badd16,Illi08,Scot92} and the first edition of 
\cite{Digg14}), there are essentially two main approaches: a Poisson process 
likelihood cross-validation approach \cite[Section~5.3]{Load99} and 
minimisation of the mean squared error in state estimation for a stationary 
isotropic Cox process \cite{Digg85}. Note that both approaches rely on a 
specific model assumption. As an alternative, we introduce a new, 
computationally simple and intuitive approach that relies solely on the 
Campbell formula \cite{Matt78}, which relates pattern averages to intensity 
function weighted spatial averages.

The plan is as follows. Section~\ref{S:preliminaries} recalls some 
basic principles of the mathematical theory of point processes and 
kernel based estimation of the intensity function. 
Section~\ref{S:bandwidth} discusses the likelihood based and state the 
estimation based methods for bandwidth selection; the new method is 
introduced in Section~\ref{S:Hamilton}. The results of a simulation 
study to assess the performance of the three methods is presented in 
Section~\ref{S:numerics} and the paper closes with a discussion of 
topics for further work.

\section{Preliminaries}
\label{S:preliminaries}

\subsection{The intensity function}

Throughout this paper, let $\Psi$ be a point process in $\R^d$, $d\geq 1$, 
observed within some non-empty open and bounded observation window 
$W \subseteq \R^d$. Hence, its realisations consist of a finite number of 
points scattered in $W$ \cite{DaleVere08,Digg14,Illi08,Lies00}. Examples 
include the arrival times of customers ($d=1$), the locations of trees in 
a forest stand ($d=2$) and the positions of galaxies in space ($d=3$). 

The abundance of points as a function of location is captured by the
{\em intensity function\/} $\lambda: \R^d \to \R^+$. 
Heuristically,
given an infinitesimal neighbourhood $dx \subseteq \R^d$ of $x\in\R^d$, of size $dx$, 
\(
\lambda(x) \, dx
\)
is the infinitesimal probability of finding a point of $\Psi$ in the region $dx$. 
More formally, writing $\Psi(A)$ for the number of points $\Psi$ places in some Borel $A\subseteq\R^d$, $\lambda$ is defined as a 
Radon--Nikodym derivative of the first order moment measure $A\mapsto\E\Psi(A)$, provided 
it exists \cite[Section~9.5]{DaleVere08}.

From now on, we will assume that $\Psi$ admits a well-defined intensity 
function $\lambda$. Then, the Campbell theorem \cite[p.~65]{DaleVere08}
states that for any measurable, non-negative function $h: \R^d \to \R^+$,
\begin{equation}
\label{Campbell}
\E\left[ \sum_{x\in\Psi} h(x) \right]
=
\int_{\R^d} h(x) \, \lambda(x) \, dx.
\end{equation}

\subsection{Poisson and Cox point process models}
\label{S:Poisson}

Let $W\neq\emptyset$ be a bounded open subset of $\R^d$. Then the {\em Poisson
process\/} with intensity function $\lambda: W \to \R^+$ on $W$ 
is constructed as follows. First, generate a Poisson distributed
random variable $N$ with rate parameter 
\[
 \int_W \lambda(x) \, dx.
\]
Then, upon the outcome $N=n$, sample $n$ independent and identically
distributed points with common probability density function
\[
\frac{\lambda(x)}{
 \int_W \lambda(y) \, dy},
 \quad x\in W.
\]
The ensemble of the points thus generated form a realisation of 
the desired Poisson process. This model is particularly amenable to 
calculations due to the strong independence assumptions. In particular, 
the log likelihood of a pattern $\x = \{ x_1, \dots, x_n \}$ reads
\begin{equation}
\sum_{i=1}^n \log \lambda(x_i) - \int_W \lambda(x) \, dx.
\label{e:loglik}
\end{equation}
The first term captures the probability of points being placed at $x_i$, $i=1,\ldots,n$, and 
the second term that of no points falling anywhere else.

A {\em Cox process\/} is the generalisation of a Poisson process that
allows for a random intensity function $\Lambda$. Such models are
appropriate to describe clustering due to latent environmental 
heterogeneity. We have that
\begin{equation}
\lambda(x) = \E \Lambda(x)
\label{e:intensCox}
\end{equation}
for every $x\in W$. The log likelihood depends on the distribution
of $\Lambda$ and is usually not available in closed form. Further
details on these and several other models can be found 
in, for example, \cite{DaleVere08,Digg14,Illi08}.

\subsection{Kernel estimation}

A {\em kernel function\/} \cite{Silv86} $\kappa: \R^d \rightarrow \R^+$ 
is defined to be a $d$-dimensional symmetric probability density function. 
Suppose that the point process $\Psi$ is observed in $W \subseteq \R^d$.
Then, given a bandwidth or scale parameter $h > 0$, the intensity function 
may be estimated by
\begin{equation}
\label{e:KernelEstimator}
\widehat \lambda(x;h)
=
\widehat \lambda(x; h, \Psi, W)
=
h^{-d} \sum_{y\in\Psi\cap W} 
\kappa\left( \frac{x - y}{h}\right) w_h(x, y)^{-1}, 
\quad x\in W,
\end{equation}
where $w_h(x, y)$ is an edge correction factor. Note that
$w_h \equiv 1$ corresponds to no edge correction. The global
edge correction factor proposed by Berman and Diggle 
\cite{BermDigg89,Digg85} is given by
\[
w_h(x, y) = h^{-d} \int_W \kappa\left( \frac{ x - u }{h} \right) du,
\]
whereas the choice of the local factor
\[
w_h(x, y) = h^{-d} \int_W \kappa\left( \frac{u-y}{h} \right) du, 
\]
suggested in \cite{Lies12}, results in the mass preservation
property that
\begin{equation}
\label{e:MassPreservation}
\int_W \widehat \lambda(x;h) \, dx = \Psi(W).
\end{equation}
At least for the
Beta kernels discussed below, which are strictly positive
on an open ball of radius $h$ around the centre point, the assumption 
that $W$ is open implies that neither edge correction factor can be zero. 
Note that upon taking expectations on both sides in \eqref{e:MassPreservation} we obtain unbiasedness.

For $\kappa$, one may for instance pick a member of the class of Beta kernels, which are also known as multiweight kernels \cite{Hall04}.
More specifically, for any $\gamma \geq 0$, set
\begin{equation}
\label{e:beta}
\kappa(x) = \kappa^\gamma(x) = 
\frac{\Gamma \left(\frac{d}{2} + \gamma + 1\right)}{
\pi^{d/2} \Gamma \left(\gamma+1\right)}
(1 - x^T x)^{\gamma} \, 1\{ x \in B(0, 1) \},
\quad x\in\R^d.
\end{equation}
Here $B(0,1)$ is the closed unit ball centred at the origin.
Specific examples include the box kernel ($\gamma=0$) and the 
Epanechnikov kernel ($\gamma=1$). Another widely used choice is
the Gaussian kernel
\begin{equation}
\label{e:Gauss}
\kappa(x) = (2\pi)^{-d/2} \exp(- x^T x / 2 ), \quad x \in \R^d.
\end{equation}
More details and further examples can be found in \cite{BowmAzza97,Silv86}. 

Note that all kernels mentioned above are isotropic in the sense 
that up to a dimension dependent constant $c_d$, $\kappa(x) = 
\kappa_d(x) \propto \kappa_1(\|x\|)$ is equal to its univariate 
counterpart evaluated at the norm $\|x\|$. A further option would 
be to define $\kappa(x)$, $x\in\R^d$, as a product of univariate kernels: $\kappa(x) = \prod_{i=1}^{d}\kappa_i(x_i)$, where each 
$\kappa_i: \R\to [0,\infty)$, $i=1,\ldots,d$, is some univariate 
kernel. 
The Gaussian kernel is an example of this approach. It is 
not a Beta kernel since it has unbounded support, but it can be seen as a degenerate limiting case upon proper scaling.

\subsection{Bias and variance}

Assume that $\Psi$ admits a second order product density
$\rho^{(2)}(x,y) \, dx dy$ which we interpret as the probability 
that $\Psi$ places points in each of two disjoint infinitesimal regions $dx$ and $dy$ around $x$ and $y$. Then, the Campbell--Mecke theorem 
\cite[Lemma~9.5.IV]{DaleVere08} implies that the first two moments of 
(\ref{e:KernelEstimator}) exist and are given by
\begin{eqnarray}
\label{e:bias}
\E \left[ \widehat \lambda(x; h) \right] 
& = & 
h^{-d} \int_W \kappa\left( \frac{x-y}{h} \right) \frac{\lambda(y)}{
 w_h(x, y) } \, dy; \\
\nonumber
\E \left[ \widehat \lambda(x; h)^2 \right] 
& = & 
h^{-2d} \int_W \int_W \kappa\left( \frac{x-y_1}{h} \right) 
\kappa\left( \frac{x-y_2}{h} \right) 
\frac{\rho^{(2)}(y_1, y_2)}{
w_h(x, y_1) w_h(x, y_2) } \, dy_1 dy_2 \\
&+&  
h^{-2d} \int_W \kappa\left( \frac{x-y}{h} \right)^2  \frac{\lambda(y)}{
w_h(x, y)^{2} } \, dy.
\label{e:error}
\end{eqnarray}
Note that, in general, the moments depend on the location $x$.

The quality of a kernel estimator may be measured by
the mean integrated squared error
\begin{eqnarray} 
MISE(\widehat\lambda(\cdot; h))
& = &
\E\left[\int_W \left(\widehat\lambda(x;h) - \lambda(x)\right)^2 dx\right]
=
\int_W \E\left[ 
   \left( \widehat\lambda(x;h) - \lambda(x) \right)^2 
\right] dx 
\nonumber
\\
&= & \int_W \left[ {\rm{Var}}(\widehat\lambda(x;h)) + 
   {\rm{bias}}^2(\widehat\lambda(x;h)) \right] dx,
\label{e:MISE}
\end{eqnarray}
where 
\(
{\rm{bias}}(\widehat\lambda(x;h)) = 
   \E\widehat\lambda(x;h) - \lambda(x).
\)
In practice, since (\ref{e:MISE}) depends on the unknown intensity and
second order product density functions, it has only limited value.

\section{Bandwidth selection}
\label{S:bandwidth}

We begin this section by reviewing two common methods to select 
a proper bandwidth. For simplicity, we illustrate the methods for
the box kernel in two dimensions.

\subsection{State estimation for isotropic Cox processes}
For the planar case, 
Diggle \cite{Digg85} suggested to treat $\widehat{\lambda}$ as a 
state estimator for the (random) intensity function $\Lambda$ of a 
stationary isotropic Cox process $\Psi$ and minimise the mean squared error
\[
\E\left[ \left( \widehat{\lambda(0;h, \Psi)} - 
                \Lambda(0) \right)^2 \right]
\]
for an arbitrary origin. For simplicity, we ignore edge correction.
Then, analogously to (\ref{e:intensCox}), the second order product density of a Cox 
process is given by $\rho^{(2)}(x,y) = \E ( \Lambda(x) \Lambda(y) )$.
Since $\Psi$ is assumed to be stationary and isotropic, 
$\rho^{(2)}(x,y) = \rho^{(2)}(||x-y||)$ is a function of the distance
$|| x - y ||$ between $x$ and $y$ only. 
Hence, with slight abuse of 
notation, the mean squared error reads \cite[Chapter 5.3]{Digg14}
\begin{equation}
\label{e:bwDiggle}
\rho^{(2)}(0) +
\frac{1}{\pi^2 h^4} \int_{B(0,h)} \int_{B(0,h)} \rho^{(2)}( || x - y || ) \, dx dy +
\frac{\lambda}{\pi h^2} \left[ 1 - 2\lambda K(h) \right],
\end{equation}
where 
\begin{equation}
\label{e:Kfunction}
 \lambda^2 K(h) =  \int_{B(0,h)} \rho^{(2)}(||x||) \, dx .
\end{equation}
It is important to note that the mean squared error is with respect to
state estimation, not estimation of the constant intensity function! 
Using the transformation $z = x-y$ and changing the order of integration,
\[
\int_{B(0,h)} \int_{B(0,h)} \frac{ \rho^{(2)}( || x - y || )}{\lambda^2} \, dx dy 
=
\int_{B(0,2h)} \int_{B(z,h)\cap B(0,h)} g(||z||) \, dx dz 
\]
\[
= \int_0^{2h} \left(2 h^2 \arccos (t/(2h)) 
- (t/2) \sqrt( 4h^2 - t^2)\right) \, dK(t),
\]
where $g(||z||)=\rho^{(2)}(||z||)/\lambda^2$ is the pair correlation function. 
Hence, the integral can be evaluated numerically based on an estimator of $K(t)$. 

From a numerical point of view, to implement this bandwidth selection
method, one needs an estimate of the constant intensity $\lambda$, an estimator $\hat K$ of
the $K$-function, a Riemann integral over the bandwidth range and an 
optimisation algorithm. The estimator $\hat K$ requires a pattern with
at least two points as well as some edge correction, which limits the 
range of $h$-values one can consider.

\subsection{Cross-validation for Poisson processes}

Likelihood cross-validation \cite{Badd16,Load99} ignores all
interaction and assumes that $\Psi$ is an inhomogeneous Poisson
process. By (\ref{e:loglik}), the leave-one-out cross-validation 
log likelihood reads
\begin{equation}
\label{e:bwPPL}
PPL_{\kappa}(h;\Psi, W)=
\sum_{x\in\Psi\cap W} \log \widehat \lambda(x; h, \Psi\setminus \{ x \}, W)
- \int_W \widehat \lambda( u; h, \Psi, W) \, du.
\end{equation}
Note that conditions have to be imposed to ensure that the function
$\widehat \lambda$ is strictly positive.

We need to specify which edge correction, if any, is used. 
In the planar case the default implementation {\tt bw.ppl} in the R-package {\tt spatstat} \cite{Badd16} uses global edge correction,
but simulations suggest the clearest optimum being found when ignoring edge effects. In the latter case, 
when employing the box kernel, 
\[
PPL_{\kappa}(h;\Psi, W)=
\sum_{x\in\Psi\cap W} \log \left( 
  \frac{ \Psi( B(x,h) \cap W) - 1 }{\pi h^2} \right)
- \frac{1}{\pi h^2} \sum_{x\in \Psi \cap W}  \ell( B(x,h) \cap W ),
\]
where $\ell(\cdot)$ denotes Lebesgue measure. 
Under global edge correction, (\ref{e:bwPPL}) reads
\[
\sum_{x\in\Psi\cap W} \log \left( \frac{ \Psi( B(x,h) \cap W ) - 1 } {
\ell( B(x,h) \cap W ) } \right) 
- \int_W \frac{ \Psi(B(u,h) \cap W ) }{ \ell(B(u,h) \cap W ) } \, du
\]
and under mass preserving local edge correction, one optimises
\[
\sum_{x\in\Psi\cap W} \log \left(
  \sum_{x\neq y \in \Psi \cap W} \frac{ 1\{ y\in B(x,h) \} }{ 
                                  \ell( B(y,h) \cap W ) } \right)
\]
over $h$. The logarithms in both cases are well defined for $h$ at least equal to the minimal interpoint distance since we assume that $W$ is open.

From a numerical point of view, to implement this bandwidth selection
method, one needs a Riemann integral over the observation window and an 
optimisation algorithm. Also, the data pattern must consist of at 
least two points. Moreover, the type of edge correction used, if any,
influences the result.

\section{A new approach}
\label{S:Hamilton}

Consider the function $h: \R^d \to \R+$ defined by $h(x) = 1\{ x \in W \} /
\lambda(x)$, which is measurable if $\lambda(x) > 0$ for $x \in W$. 
Applying the Campbell formula (\ref{Campbell}) to $h$ we obtain
\bea
\label{HamiltonPrinciple}
\E\left[ \sum_{x\in\Psi\cap W} \frac{1}{\lambda(x)} \right] 
=
\int_{W}\frac{1}{\lambda(x)} \lambda(x) \, dx
=
\ell(W).
\eea 
In other words, $\sum_{x\in\Psi\cap W} \lambda(x)^{-1}$ is an unbiased estimator 
of the window size $\ell(W)$. If $\lambda(\cdot)$ is replaced by its estimated 
counterpart, $\widehat\lambda(\cdot)$, the left hand side of 
(\ref{HamiltonPrinciple}) is a function of the bandwidth while the right 
hand side is not. We may therefore minimise the discrepancy between 
$\ell(W)$ and $\sum_{x\in\Psi\cap W} \widehat\lambda(x; h, \Psi, W)^{-1}$ 
to select an appropriate $h$. 
Formally, we define 
\beann
\label{e:HamFun}
T_{\kappa}(h;\Psi, W)
=
\left\{
\begin{array}{ll}
\displaystyle
\sum_{x\in\Psi\cap W} \frac{1}{\widehat\lambda(x;h, \Psi, W)}
&\text{if } \Psi\cap W \neq \emptyset, \\
\ell(W)
&
\text{otherwise},
\end{array}
\right.
\eeann
and choose bandwidth $h>0$ by minimising 
\begin{equation}
\label{e:DefHam}
F_{\kappa}(h;\Psi, W, \ell(W))
=
\left(
T_{\kappa}(h;\Psi, W)
-\ell(W)
\right)^2.
\end{equation}
Here $\widehat\lambda(x;h, \Psi, W)$ is given by (\ref{e:KernelEstimator}).

Note that since $W$ is bounded, $\Psi\cap W$ almost surely contains finitely 
many points. We use the convention that when $\Psi\cap W=\emptyset$, 
then $T_{\kappa}(h;\Psi, W)\equiv\ell(W)$.
This is simply a way of saying that if nothing is observed, there is 
nothing to estimate, so we always estimate correctly. 

\subsection{Properties}

Turning to the properties of $T_{\kappa}(h;\Psi, W)$, by 
looking closer at the structure of $T_{\kappa}(h;\Psi, W)$, we note that 
when $\Psi\cap W\neq\emptyset$, 
\beann
T_{\kappa}(h;\Psi, W)
&=&\frac{1}{\prod_{x\in\Psi\cap W} \widehat\lambda(x;h, \Psi, W)}
\sum_{z\in\Psi\cap W}
\prod_{x\in\Psi\cap W\setminus\{z\}}\widehat\lambda(x;h, \Psi, W)
,
\eeann
i.e.\ the structure is that of a `leave-one-out' kind: we compare 
$\prod_{x\in\Psi\cap W}\widehat\lambda(x;h)$ to all versions where we exclude 
one of its terms. Thus $T_{\kappa}(h;\Psi, W)$ is more complex than one 
might initially anticipate.

Next, we consider the continuity properties of $T_\kappa(\cdot; \Psi, W)$ 
and its limits as the bandwidth approaches zero and infinity. The proof 
can be found in the appendix.

\begin{lemma} \label{LemmaConvergence}
Let $\Psi$ be a point process in $\R^d$, observed in some non-empty
open and bounded window $W$, and exclude the trivial case that 
$\Psi \cap W = \emptyset$. Let $\kappa(\cdot)$ be a Gaussian or Beta 
kernel with $\gamma>0$. Then $T_{\kappa}(h; \Psi, W)$ is a continuous 
function of $h$. For the box kernel, $T_{\kappa}(h;\Psi, W)$ is piecewise 
continuous in $h$. In all cases, 
\[
\lim_{h\to 0} T_{\kappa}(h;\Psi, W) =0.
\]
Also,  
\[
\lim_{h\to \infty} T_{\kappa}(h;\Psi, W) = \infty
\]
when $w_h \equiv 1$ and 
\[
\lim_{h\to \infty} T_{\kappa}(h;\Psi, W) = \ell(W)
\]
for the global and local edge corrections. 
\end{lemma}

It is important to note that Lemma~\ref{LemmaConvergence} may not 
hold if a leave-one-out estimator is used for the intensity function. 
The above lemma demonstrates that when $w_h(\cdot)\equiv1$, with 
probability 1, the minimum of (\ref{e:DefHam}) is zero and that this 
minimum is attained. It need not be unique since
$T_{\kappa}(h;\Psi, W)$ is not necessarily a monotone function of $h$. We do, however, conjecture that monotonicity holds when $\kappa$ is the Gaussian kernel.

\section{Numerical evaluation}
\label{S:numerics}

To compare the three bandwidth selection approaches outlined in
Sections~\ref{S:bandwidth}--\ref{S:Hamilton}, we carry out a simulation
study. The point process models we consider have been selected so as 
to allow explicit formulas for their intensity functions. 
Given a set of parameters, we generate 
100 simulations of each model, 
on the window $[0,1]^2$. 
For each of the three bandwidth selection approaches considered, we estimate the bandwidth using no edge correction, with a discretisation of $128$ values in the range $[0.01, 1.5]$ and, for the cross-validation method, 
we use a spatial discretisation of $[0,1]^2$ in a $128\times 128$ grid for the numerical evaluation of the integral in \eqref{e:bwPPL}. 
To assess the quality of the selection approaches by means of \eqref{e:MISE}, the average integrated squared error over the $100$ samples was calculated for each method, using a Gaussian kernel with the selected bandwidths and applying local edge correction. To express the results in a comparable scale, we divide them by the expected number of points in $[0,1]^2$. 
Calculations were carried out using the R-package {\tt spatstat} \cite{Badd16} and below we give our conclusions together with examples of the experiments described above. 

\subsection{Poisson processes}
We start by evaluating a set of Poisson processes (see Section~\ref{S:Poisson}), with different intensity functions. The results below correspond to one round of experiments and the conclusions are based on our overall observations.

For all cases, in the high level intensity setting, the state estimation approach performs the best and the new approach has the highest average integrated squared error. 
In the homogeneous case, we see that for the low intensity level the likelihood based approach is performing the best 
and the state estimation approach is giving rise to the highest average integrated squared error. For the medium level intensity, the new and the state estimation approaches yield comparable average integrated squared errors, 
both being outperformed by the likelihood based approach. 
In the inhomogeneous cases, for small and medium intensities, it seems that the likelihood based approach has the best performance, followed by the new approach. 

\subsubsection{Homogeneous Poisson process}

In the first experiment, we generated $100$ independent realisations 
of a homogeneous Poisson process in the unit square for low, medium and high intensity values 
($\lambda = 10, 50, 250$). The expected number of points in $[0,1]^2$ is $\lambda$.
The results are summarised in Table~\ref{T:PoissonHom}.

\begin{table}[hbt]
\begin{tabular}{l|rrr}
& New & State & Likelihood \\
\hline
$\lambda = 10$ & 3.4 & 11.0 & 2.8 \\
$\lambda = 50$ & 10.0 & 10.6 & 5.9 \\
$\lambda = 250$ & 30.7 & 12.2 & 16.4 \\
\end{tabular}
\caption{Average integrated squared error, divided by the expected number of points, over $100$ simulations of a
homogeneous Poisson process on the unit square.}
\label{T:PoissonHom}
\end{table}

\subsubsection{Poisson process with linear trend}

In the next experiment, we generated $100$ independent realisations 
of a Poisson process in the unit square with intensity function
\[
\lambda(x,y) = 10 + \alpha x, \quad (x,y) \in [0,1]^2,
\]
for weak, medium and strong trend ($\alpha = 1, 80, 480$). 
The expected number of points in $[0,1]^2$ is $10 + \alpha/2$.
The results are summarised in Table~\ref{T:PoissonTrend}.

\begin{table}[hbt]
\begin{tabular}{l|rrr}
& New & State & Likelihood \\
\hline
$\alpha = 1$ & 4.2 & 13.0 & 3.8 \\
$\alpha = 80$ & 8.9 & 11.4 & 7.4 \\
$\alpha = 480$ & 23.0 & 14.8 & 17.9 \\
\end{tabular}
\caption{Average integrated squared error, divided by the expected number of points, over $100$ simulations of a
Poisson process with linear trend on the unit square.}
\label{T:PoissonTrend}
\end{table}

\subsubsection{Poisson process with modulation}

Finally, we generated $100$ independent realisations 
of a Poisson process in the unit square with intensity function
\[
\lambda(x,y) = \alpha + \beta \cos(10 x), \quad (x,y) \in [0,1]^2.
\]
To generate patterns with a low, medium and large intensity, we
considered the $(\alpha, \beta)$ combinations $(10, 2)$, $(50, 20)$
and $(250, 100)$. 
The expected number of points in $[0,1]^2$ is $\alpha+\beta\sin(10)/10$.
The results are summarised in Table~\ref{T:PoissonWave}.

\begin{table}[hbt]
\begin{tabular}{l|rrr}
& New & State & Likelihood \\
\hline
$(\alpha, \beta) = (10, 2)$ & 4.0 & 11.8 & 3.7\\
$(\alpha, \beta) = (50, 20)$ & 10.8 & 12.7 & 9.0\\
$(\alpha, \beta) = (250, 100)$ & 29.6 & 17.9 & 20.0\\
\end{tabular}
\caption{Average integrated squared error, divided by the expected number of points, over $100$ simulations of a
modulated Poisson process on the unit square.}
\label{T:PoissonWave}
\end{table}

\subsection{Cox processes}

We next turn to a class of clustered point processes, namely Cox processes (see Section~\ref{S:Poisson}). 
As indicated by the specific experiments below, for each model type considered, the new approach seems to strongly outperform the competing approaches.

\subsubsection{Homogeneous Mat\'ern cluster process}

In the first experiment we generated $100$ independent realisations of a homogeneous Mat\'ern cluster process \cite{SKM} in the unit square, for various
degrees of clustering. This is a Cox process with
\[
\Lambda(y) =  \frac{1}{\pi r^2} \sum_{x\in\Phi} 1\{ y \in B(x,r) \},
\]
where $\Phi$ is a homogeneous Poisson process with intensity $\kappa$ and $B(x,r)$ is the closed ball of radius $r$, centred at $x\in \R^2$.
In words, each point of $\Phi$ acts as a parent to a family that
consists of a Poisson number of offspring (mean size $\mu$), born independently of each other at positions that are uniformly 
distributed in a ball of radius $r$ around the parent. The Mat\'ern
cluster process $\Psi$ consists of the ensemble of all offspring.

We considered two levels, $\kappa=10, 20$, 
for the parent intensity, two ranges of clustering, $r=0.05, 0.1$, and two mean offspring sizes, $\mu = 3, 10$. To avoid edge effects,
the parent process was generated on a dilated window. Note that
the resulting point process is homogeneous with intensity 
$\kappa \mu$. As an aside, for general parent intensity 
functions $\lambda(\cdot)$, the resulting point process 
has intensity function
\[
\lambda(x) = \frac{\mu}{\pi r^2} \int_{B(x,y)} \lambda(u) \, du.
\]

The results are summarised in Table~\ref{T:ClusterHom} and we note that the expected number of points in $[0,1]^2$ is $\kappa\mu$. 
Throughout, the new approach outperforms its competitors.  For a low degree of clustering (small values of $\mu$), 
the likelihood based method outperforms the state estimation approach; for larger degrees the state estimation method works better. 
The state estimation and likelihood based methods tend to have decreased integrated squared errors when clusters are more diffuse.

\begin{table}[hbt]
\begin{tabular}{l|rrr}
& New & State & Likelihood \\
\hline
$(\kappa, r, \mu) = (10, 0.05, 3) $ 
&19.5 & 233.5 & 158.3\\
$(\kappa, r, \mu) = (10, 0.1, 3) $ 
&15.0 & 64.5 & 47.7\\
$(\kappa, r, \mu) = (10, 0.05, 10) $ 
&44.5 & 627.4 & 783.9\\
$(\kappa, r, \mu) = (10, 0.1, 10) $ 
&44.2 & 175.2 & 214.6\\
$(\kappa, r, \mu) = (20, 0.05, 3) $ 
&22.8 & 203.8 & 158.4\\
$(\kappa, r, \mu) = (20, 0.1, 3) $ 
& 21.7 & 58.5 & 50.6\\
$(\kappa, r, \mu) = (20, 0.05, 10) $ 
&62.9 & 552.5 & 765.5\\
$(\kappa, r, \mu) = (20, 0.1, 10) $ 
&71.9 & 168.9 & 236.2\\
\end{tabular}
\caption{Average integrated squared error, divided by the expected number of points, over $100$ simulations of a
homogeneous Mat\'ern cluster process on the unit square.}
\label{T:ClusterHom}
\end{table}

\subsubsection{Homogeneous Log-Gaussian Cox process}
\label{S:HomLGCP}

In the next experiment, we generated $100$ independent realisations
of a homogeneous Cox process as follows. 
Let $Z$ be a Gaussian random field with mean zero and covariance function
\[
\sigma^2 \exp\left( - \beta || (x_1, y_1)  - (x_2, y_2) || \right)
.
\]
Consider further the random measure defined by its intensity
\[
\lambda \exp( Z(x,y) ).
\]
Then the intensity function of the resulting Cox process, a log-Gaussian Cox process \cite{Digg14,Moller98}, is given by 
\[
\lambda \exp( \sigma^2 / 2 ).
\]

We considered two levels $\lambda = 10, 50$, two degrees of variability, $\sigma^2 = 2\log(5)$ and $\sigma^2 = 2\log(2)$, as well as two degrees of clustering $\beta = 10, 50$. 
Note that the expected number of points in $[0,1]^2$ is $\lambda \exp( \sigma^2 / 2 )$. 

The results are summarised in Table~\ref{T:LGCPHom}.
As already mentioned, we see that the new approach works the best. It further seems that the likelihood based approach has the second best performance. Increasing $\beta$, that is, decreasing the range of interaction, tends to lead to a smaller integrated squared error. Increasing the variability, and hence the intensity, yields a higher integrated squared error.

\begin{table}[hbt]
\begin{tabular}{l|rrr}
& New & State & Likelihood \\
\hline
$(\lambda, \sigma^2, \beta) = (10, 2\log(5), 50)$ & 14.1 & 70.8 & 17.8\\
$(\lambda, \sigma^2, \beta) = (10, 2\log(2), 10)$ & 9.7 & 28.1 & 12.4\\
$(\lambda, \sigma^2, \beta) = (10, 2\log(5), 10)$ & 55.0 & 376.5 & 208.6\\
$(\lambda, \sigma^2, \beta) = (50, 2\log(5), 50)$ & 73.7 & 641.7 & 336.6\\
$(\lambda, \sigma^2, \beta) = (50, 2\log(2), 10)$ & 61.1 & 130.9 & 102.2\\
$(\lambda, \sigma^2, \beta) = (50, 2\log(5), 10)$ & 383.4 & 2,875.1 & 2,229.9\\
\end{tabular}
\caption{Average integrated squared error, divided by the expected number of points, over $100$ simulations of a
homogeneous log-Gaussian Cox process.}
\label{T:LGCPHom}
\end{table}

\subsubsection{Log-Gaussian Cox process with linear trend}
\label{S:LinearTrendLGCP}

Turning to the more important scenario of inhomogeneity in combination with clustering, we generated $100$ independent realisations of a Cox process as follows. Let $Z$ be the same Gaussian random field as in Section \ref{S:HomLGCP} and consider the random measure defined by its intensity
\[
\lambda(x,y) \exp( Z(x,y) ),
\]
where $\lambda$ is strictly positive. 
Then the intensity function of the resulting Cox process is
\[
\lambda(x,y) \exp( \sigma^2 / 2 ).
\]
We chose the linear trend function
\[
\lambda(x,y) = 10 + 80x,
\]
two degrees of variability, $\sigma^2 = 2\log(5)$ and 
$\sigma^2 = 2\log(2)$, as well as two degrees of clustering, $\beta = 10, 50$. 
Here the expected number of points in $[0,1]^2$ is $50 \exp( \sigma^2 / 2 )$. 

The results are summarised in Table~\ref{T:LGCPLinear} and the conclusions, in terms of performance, are the same as for the homogeneous log-Gaussian Cox process in Section \ref{S:HomLGCP}.

\begin{table}[hbt]
\begin{tabular}{l|rrr}
& New & State & Likelihood \\
\hline
$(\sigma^2, \beta) = (2\log(5), 50)$ & 
89.6 & 1477.2 & 536.0\\
$(\sigma^2, \beta) = (2\log(2), 10)$ & 
57.5 & 136.9 & 112.6\\
$(\sigma^2, \beta) = (2\log(5), 10)$ & 
335.3 & 2,960.6 & 2,251.2\\
\end{tabular}
\caption{Average integrated squared error, divided by the expected number of points, over $100$ simulations of a
log-Gaussian Cox process with linear trend.}
\label{T:LGCPLinear}
\end{table}

\subsubsection{Log-Gaussian Cox process with modulation}

In the next experiment, we generated $100$ independent realisations
of a Cox process in the same way as in Section \ref{S:LinearTrendLGCP}, but with 
\[
\lambda(x,y) = 10 + 2 \cos(10x).
\]
Here we used two degrees of variability, $\sigma^2 = 2\log(5)$ and 
$\sigma^2 = 2\log(2)$, as well as two degrees of clustering, 
$\beta = 10, 50$. 
The expected number of points in $[0,1]^2$ is $(10 + \sin(10)/5) \exp( \sigma^2 / 2 )$.

The results are summarised in Table~\ref{T:LGCPWave}. Also here the conclusions are the same. The selection \eqref{e:DefHam} is performing the best and the state estimation approach the worst, with increasing
$\beta$ leading to a smaller integrated squared error and increased variability yielding a higher integrated squared
error. 

\begin{table}[hbt]
\begin{tabular}{l|rrr}
& New & State & Likelihood \\
\hline
$(\sigma^2, \beta) = (2\log(5), 50)$ & 
16.3 & 93.6 & 21.2\\
$(\sigma^2, \beta) = (2\log(2), 10)$ & 
9.1 & 29.4 & 11.3\\
$(\sigma^2, \beta) = (2\log(5), 10)$ & 
78.4 & 730.7 & 392.9\\
\end{tabular}
\caption{Average integrated squared error, divided by the expected number of points, over $100$ simulations of a
modulated log-Gaussian Cox process.}
\label{T:LGCPWave}
\end{table}

\subsection{Determinantal point processes}
We finally turn to planar determinantal point processes (see e.g.\ \cite{Lav15}). Such processes exhibit inhibition/regularity and allow explicit expressions for the product densities 
\[
\rho^{(n)}((x_1,y_1), \dots, (x_n,y_n)) = {\rm{det}}( C((x_i,y_i), (x_j,y_j)) )_{i,j},
\quad (x_i,y_i) \in \R^2,
\]
in terms of a kernel $C$. Here ${\rm{det}}$ denotes the determinant operator. Hence, the intensity function is given by $\lambda(x,y) = \rho^{(1)}(x,y) = C( (x,y), (x,y) )$. 
Throughout, we will use 
\[
C((x_1,y_1), (x_2,y_2)) = \sigma^2 \exp(- \beta \| (x_1,y_1) - (x_2,y_2) \| )
\]
as basis, resulting in a homogeneous determinantal point process with intensity 
\[
\lambda(x,y) = C( (x,y), (x,y) ) = \sigma^2.
\]
Hence, the expected number of points in $[0,1]^2$ is $\sigma^2$. 

\subsubsection{Homogeneous determinantal point process}
We generated $100$ independent realisations of the homogeneous model above, on the unit square, for low, medium
and high intensity values ($\sigma^2 = 10, 50, 250$). We further considered two values for $\beta$, namely $20$ and $50$. For $\sigma^2 =250$, the model is not valid for the larger of the interaction ranges, that is, for $\beta = 20$. To alleviate edge effects, we considered realisations on a dilated window of size $2/\beta$, which was restricted to the unit square. 

The results, which are summarised in Table~\ref{T:dppHom}, indicate that for small intensities the likelihood based method performs the best and the state estimation approach generates the highest average integrated squared error. 
For medium intensities the likelihood based approach performs the best, with the other two approaches giving rise to similar average integrated squared errors.
In the case of high intensities, the state estimation method performs the best and the new method gives rise to the highest average integrated squared error. 

\begin{table}[hbt]
\begin{tabular}{l|rrr}
& New & State & Likelihood \\
\hline
$(\sigma^2, \beta) = (10, 50)$ & 4.0 & 11.0 & 3.2 \\
$(\sigma^2, \beta) = (10, 20)$ & 4.1 & 13.1 & 3.2 \\
$(\sigma^2, \beta) = (50, 50)$ & 10.9 & 10.6 & 6.3 \\
$(\sigma^2, \beta) = (50, 20)$ & 9.6 & 8.4 & 4.8 \\
$(\sigma^2, \beta) = (250, 50)$ & 27.6 & 10.4 & 11.7 \\
\end{tabular}
\caption{Average integrated squared error, divided by the expected number of points, over $100$ simulations of a
homogeneous determinantal point process.}
\label{T:dppHom}
\end{table}

\subsubsection{Determinantal point process with linear trend}

Next, we applied independent thinning (see e.g.\ \cite{SKM}) to the realisations of 
the homogeneous determinantal point process described above. We employed the retention probabilities
\[
p(x,y) = \frac{10 + 80x}{90}, \quad (x,y) \in [0,1]^2,
\]
which results in the intensity being given by $\lambda(x,y) = \sigma^2p(x,y)$. 
We considered the cases $\lambda=50, 250$ and $\beta=20, 50$. The results are summarised in Table~\ref{T:dppLinear}. Here, the expected number of points in $[0,1]^2$ is $5\sigma^2/9$. 
We found indications that in the low and medium intensity cases the likelihood based approach performs the best, followed by the new approach. On the other hand, in the high intensity setting the likelihood based approach and the state estimation approach seem to perform almost equally well.

\begin{table}[hbt]
\begin{tabular}{l|rrr}
& New & State & Likelihood \\
\hline
$(\sigma^2, \beta) = (50, 50)$ & 
7.0 & 10.6 & 5.3\\
$(\sigma^2, \beta) = (50, 20)$ & 
6.2 & 9.7 & 5.1\\
$(\sigma^2, \beta) = (250, 50)$ & 
16.7 & 10.8 & 10.9\\
\end{tabular}
\caption{Average integrated squared error, divided by the expected number of points, over $100$ simulations of a
determinantal point process with linear trend.}
\label{T:dppLinear}
\end{table}

\subsubsection{Determinantal point process with modulation}

In the last experiment, we applied independent thinning to realisations of the homogeneous determinantal point process described in the previous 
section, using retention probabilities
\[
p(x,y) = \frac{ 10 + 2 \cos(10x) }{12},  \quad (x,y) \in [0,1]^2,
\]
for $\lambda=10, 50$ and $\beta=20, 50$. 
Note that the expected number of points in $[0,1]^2$ is $\sigma^2(10 + \sin(10)/5)/12$.
The results are summarised in Table~\ref{T:dppLinear}. For the smallest intensity, the new approach and the likelihood based approach seem to have similar, but also the best, performance. As we increase the intensity, the new approach tends to generate a higher average integrated squared error than the likelihood based approach. 
The state estimation method tends to perform poorly for small intensities; there may not be enough points to reliably estimate the $K$-function in \eqref{e:bwDiggle} and \eqref{e:Kfunction}.

\begin{table}[hbt]
\begin{tabular}{l|rrr}
& New & State & Likelihood \\
\hline
$(\sigma^2, \beta) = (10, 50)$ & 
4.0 & 11.7 & 4.9 \\
$(\sigma^2, \beta) = (10, 20)$ & 
3.7 & 13.2 & 3.3\\
$(\sigma^2, \beta) = (50, 50)$ & 
9.8 & 10.6 & 6.4\\
$(\sigma^2, \beta) = (50, 20)$ & 
8.5 & 8.4 & 4.8\\
\end{tabular}
\caption{Average integrated squared error, divided by the expected number of points, over $100$ simulations of a
modulated determinantal point process.}
\label{T:dppWave}
\end{table}

\section{Discussion}

We have proposed a new approach for the selection of bandwidth $h>0$ in a kernel estimator $\widehat\lambda(x;h, \Psi, W)$ of the intensity function $\lambda(x)>0$, $x\in W$, of a point process $\Psi$. The basic idea is to minimise the discrepancy between the size of the study region, $\ell(W)$, and the sum of reciprocals $T_{\kappa}(h;\Psi, W)=\sum_{x\in\Psi\cap W} \widehat\lambda(x;h, \Psi, W)^{-1}$. 
Our approach is mandated by the fact that replacing $\widehat\lambda(x;h, \Psi, W)$ by $\lambda(x)$ in $T_{\kappa}(h;\Psi, W)$ turns this statistic into an unbiased estimator of $\ell(W)$. In essence, we have transformed the problem of selecting the unknown optimal bandwidth to that of estimating the known size of the study region and, as a by-product, we obtain an estimate of the optimal bandwidth. 
Our approach, being based solely on the general Campbell theorem \eqref{Campbell}, is fully non-model based, as opposed to the current state of the art methods. These approaches, the Poisson process likelihood cross-validation approach \cite{Badd16,Load99} and the state estimation approach \cite{Digg85}, both assume that the underlying point process $\Psi$ is of a given model class.

The results of the simulation study carried out in this paper suggest the following tentative conclusions. For clustered patterns, the new method appears to be the best choice. For Poisson processes, at least for moderately sized patterns, the likelihood based cross-validation method seems to give the best results. The picture is a bit more varied for regular patterns. Broadly speaking, both the new method and the likelihood based method give good results for low and moderate intensity values; the state estimation method
seems best for dense patterns. Indeed, there are indications that the state estimation approach outperforms the other methods when the intensity
becomes very large. Finally, in terms of computational speed, the state estimation approach is the fastest of the methods considered in this paper and the Poisson likelihood cross-validation approach is the slowest.

Taking these observations into consideration, we tentatively suggest using the new approach, unless i) the pattern exhibits clear inhibition/regularity, or ii) the number of points is very large and the pattern clearly exhibits no clustering. 
When i) holds we suggest employing the likelihood based approach and when ii) holds one should consider the state estimation approach. 

The method proposed in Section~\ref{S:Hamilton} assumes that the intensity
function is strictly positive. However, it is not hard to deal with any
zero or near-zero intensity regions that may arise in practice by a simple
superposition. 
More specifically, consider a further point process $\Xi\subseteq W$ that
is independent of $\Psi$, with known intensity function $\lambda_{\Xi}(x)>0$.
Then the superposition $\Psi\cup\Xi$ has intensity \cite[p.\ 165]{SKM}
\[
\lambda_{\Psi\cup\Xi}(x) = \lambda(x) + \lambda_{\Xi}(x),
\quad x\in W,
\]
which is strictly positive. By employing \eqref{e:DefHam} to obtain an estimate $\widehat\lambda_{\Psi\cup\Xi}(\cdot;h)$ based on $(\Psi\cup\Xi)\cap W$, we may subtract the known intensity $\lambda_{\Xi}(\cdot)$ to obtain the final estimate $\widehat\lambda(\cdot;h)$. 
Some care must be taken when choosing $\Xi$. If $\lambda_{\Xi}$ is too small,
the superposition will not help in solving the numerical problem of having
near-zero intensities. However, if $\lambda_{\Xi}$ is too large, 
the features of $\Psi$ may not be detectable through $\Psi\cup\Xi$. 
Note that in practise the execution amounts to repeating this procedure for a large number of simulations of $\Xi$.

It may be noted that \eqref{e:DefHam} can be expressed as a general intensity estimation criterion. More specifically, given some collection $\Gamma$ of functions $\gamma:E\to(0,\infty)$, which are estimating the underlying intensity $\lambda(\cdot)$, 
one would use as estimate $\widehat\lambda(\cdot)$ a minimiser of 
$F(\gamma)=(\sum_{x\in\Psi\cap W} \gamma(x)^{-1}-\ell(W))^2$, $\gamma\in\Gamma$. 
In fact, this is a way of estimating marginal Radon-Nikodym derivatives, without explicit knowledge of the multivariate structures. This bring us to our ongoing work, where we explore, among other things, estimation of higher order product densities, adaptive kernel estimation, intensity estimation for marked and/or spatio-temporal point processes, as well as parametric estimation.

\section*{Appendix: Proof of Lemma \ref{LemmaConvergence}}

First note that the $\kappa^\gamma$, as defined in (\ref{e:beta}), are 
continuous functions for $\gamma > 0$, as is the Gaussian kernel 
(\ref{e:Gauss}). The function $h\to 1/h$ is continuous on the open 
interval $(0,\infty)$, whereby the composition $h\to \kappa^\gamma(x/h)$ 
is also continuous in $h$ on $(0,\infty)$. 
As for the local and global edge correction functions, by the dominated 
convergence theorem, using the fact that $W$ and $\kappa^\gamma$ are
bounded, also $w_h(x,y)$ is continuous in $h$; moreover, it takes strictly 
positive values since $W$ is open. Therefore, $\widehat \lambda(x; h)$ is 
also continuous in $h$ on $(0,\infty)$. Finally, since for $x\in \Psi\cap W$, 
the kernel estimator $\widehat \lambda(x; h) \geq 
h^{-d} \kappa^\gamma(0) / w_h(x, x)  > 0$,  we reach the conclusion that
$T_{\kappa^\gamma}(h; \Psi, W)$ is continuous in $h\in (0, \infty)$.  
Box kernels are discontinuous at $1$, making $T_{\kappa^0}$ piecewise
continuous.

Now let $h$ tend to zero. For the Beta kernels with $\gamma \geq 0$, 
the support of $h^{-d} \kappa^\gamma( (\cdot - y) / h)$ is the ball centred 
at $y$ with radius $h$. Since the volume of such a ball tends to zero as 
$h$ tends to zero and $W$ is open, using the fact that $\kappa^\gamma$ is a 
symmetric probability density, we obtain that 
$\lim_{h\to 0} w_h(x,y) = 1$ for all $x,y\in W$ for global and local edge correction. This holds trivially 
if no edge correction is applied. For the Gaussian kernel, note that $h^{-d} \kappa( (\cdot - y) /h )$ corresponds to the probability density of $d$ independent normally distributed random variables
with standard deviation $h$ centred at the components of $y/h$. 
Take a sequence $h_n \to 0$ and write $X_n$ for the corresponding
Gaussian random vector. Then $X_n$ converges in distribution to 
the Dirac mass at $y$ by L\'evy's continuity theorem. 
Since $W$ is open, it is a continuity set, and $w_h$ tends to $1$
for local edge correction. The global case follows from the
symmetry of $\kappa$ and the case $w_h \equiv 1$ is trivial.
Note that for all considered kernels, $0 \leq \kappa(\cdot) \leq \kappa(0)$. 
For $x=y$ and all $h>0$, $\kappa( (x-y)/h ) = \kappa(0)$. Therefore, 
having excluded the pathological case that $\Psi\cap W$ is empty, 
$\sum_{y\in \Psi \cap W} \kappa((x-y)/h) w_h^{-1}(x,y) \geq 
\kappa(0) w_h^{-1}(x,x)$ which tends to $\kappa(0)> 0$ as $h \to 0$. 
Observing that $h^d$ tends to zero with $h$, we obtain the desired limit
$\lim_{h\to 0} T_{\kappa}(h;\Psi, W) = 0$. 

Next, let $h$ tend to infinity. Since $\kappa(\cdot) \leq \kappa(0)$,
$\lim_{h\to\infty} h^{-d} \kappa( (x-y)/h) = 0$ for all $x,y\in W$. Therefore, 
$T_{\kappa}(h; \Psi, W)$ tends to infinity when $w_h\equiv 1$ and $\Psi\cap 
W \neq \emptyset$. 
If one does correct for edge effects, by the continuity
of $\kappa$ in $0$ for all considered kernels, $\kappa((x-y)/h)$ tends
to $\kappa(0)$ as $h\to\infty$. Furthermore, since $W$ is bounded, by the dominated convergence 
theorem, 
\[
 \int_W \kappa( (x-u)/h ) \, du \to \kappa(0) \, \ell(W).
\]
Upon using the symmetry of $\kappa$, we obtain that for both local and
global edge correction, $h^d w_h(x,y) \to \kappa(0) \ell(W)$. Therefore
$\widehat \lambda(x; h) \to |\Psi \cap W| / \ell(W)$ and 
$\lim_{h\to 0} T_{\kappa}(h;\Psi, W) = \ell(W)$. 
This completes the proof.

\end{document}